\documentstyle[12pt,epsf]{article}

\author{S.~A.~Abel$^a$ and C.~A.~Savoy$^{b}$\\
{\small $^a$Theory Division, Cern 1211, Geneva 23, Switzerland}\\
{\small $^b$CEA-SACLAY, Service de Physique Th\'eorique,}\\
{\small F-91191 Gif-sur-Yvette Cedex, France}    }

\title{On Metastability in Supersymmetric Models} 
\newcommand{\hepph}[1]{{\tt hep-ph/#1 }}
\newcommand{\phrd}[3]{{{\it Phys.~Rev.}~{\bf D#1} (#3) #2}}
\newcommand{\plb}[3]{{{\it Phys.~Lett.}~{\bf B#1} (#3) #2}}
\newcommand{\prd}[3]{{{\it Phys.~Rev.}~{\bf D#1} (#3) #2}}
\newcommand{\prl}[3]{{{\it Phys.~Rev.~Lett}~{\bf #1} (#3) #2}}
\newcommand{\npb}[3]{{{\it Nucl.~Phys.}~{\bf B#1} (#3) #2}}

\newcommand{\rpp}[3]{{{\it Rept.~Prog.~Phys.}~{\bf #1} (#3) #2}}
\newcommand{\leqsim}{\,\raisebox{-0.6ex}{$\buildrel < \over \sim$}\,}
\newcommand{\geqsim}{\,\raisebox{-0.6ex}{$\buildrel > \over \sim$}\,}
\newcommand{\be}{\begin{equation}}
\newcommand{\ee}{\end{equation}}
\newcommand{\ba}{\begin{eqnarray}}
\newcommand{\ea}{\end{eqnarray}}
\newcommand{\etal}{\mbox{\em et al}}
\newcommand{\ie}{\mbox{\em i.e.~}}
\newcommand{\eg}{\mbox{\em e.g.~}}
\newcommand{\cf}{\mbox{\em c.f.~}}
\newcommand{\nn}{\nonumber}
\newcommand{\dif}{\mbox{d}}
\def\gev{\,{\rm GeV}}

\newcommand{\smallfrac}[2]{\frac{\mbox{\small #1}}{\mbox{\small #2}}}

\begin{document} 

\maketitle


\begin{abstract} 
\noindent
We make a critical reappraisal of `unbounded-from-below' (UFB)
constraints in the MSSM and $R$-parity violating models.  We explain
why the `traditional' UFB bounds are neither necessary nor sufficient
and propose, instead, a sufficient condition which ensures that there
are no {\em local} minima along the flat directions.  This conservative
(but meaningful) condition divides the parameter space into regions
which are allowed, regardless of cosmology, and regions in which
cosmology is expected to play a major role.  We study both conditions
at low $\tan\beta $ and obtain analytic approximations to the UFB
bounds for all low $\tan\beta$ ($<15 $).  Finally we show that
$R$-parity violation just below current experimental limits avoids UFB
problems by lifting the dangerous flat directions.
 
\end{abstract} 
\vspace{0.5 cm}
\newpage
\section{Introduction}

Metastability is an important issue for supersymmetric models because
of the existence of directions which are flat in the limit of unbroken
supersymmetry. Supersymmetry breaking can result in the physical vacuum
being metastable; an uncomfortable situation which some authors believe
is grounds for a model to be rejected. Others argue that a model which
is metastable is fine, provided that the vacuum could not have decayed
within the lifetime of the universe (see Refs.\cite{ccb1}-\cite{riotto}
for examples of both).  This particular question is unlikely to be
resolved until more is known about early cosmology. In the meantime, it
is important to ask whether metastability is actually the best
criterion for deciding if a flat direction is going to be troublesome
for cosmology.  In fact in this paper we will argue that it is not. We
shall also be considering a non-cosmological solution to the problems
posed by flat directions -- $R$-parity violation.

To address these issues we shall re-examine $F$ and $D$-flat directions
in the Minimal Supersymmetric Standard Model (MSSM) and in $R$-parity
violating models. $D$-flat directions can be driven negative by the
soft supersymmetry breaking terms leading to undesirable charge and
colour breaking minima~\cite{ccb1,komatsu,ccb,casas}.  When the
direction is also $F$-flat, particularly strong constraints on
parameter space arise which go by the misnomer of `Unbounded From
Below' (UFB) constraints~\cite{komatsu,ccb,casas}. Provided that all
the supersymmetry breaking mass-squared terms are positive at the high
(we shall assume GUT) scale, what actually happens is that a minimum
forms radiatively at a scale of order $few\times m_W/h_b$.  Imposing
the condition that this minimum is not the global one results in what
we shall refer to as the `traditional' UFB bound.

Following the original work on charge and colour breaking
vacua~\cite{ccb1,komatsu}, such bounds have been analysed directly
using numerical runnings of the MSSM~\cite{ccb,casas}.  Although it
gives good accuracy, this type of analysis is less than transparent.
Instead, we will work with analytic solutions to the one-loop RGEs in
the approximation that the bottom and tau contributions are negligible
(valid for all low $\tan\beta $).  These solutions will enable us to
derive traditional UFB limits (on $m_0$) for all low $\tan\beta $ (as a
function of $M_{1/2}$, $\tan\beta$ and and $A_0$). Thus by sacrificing
some accuracy we gain a lot of flexibility and also some insight into
how the potentials along the flat directions behave as we change the
parameters.  (We should stress that we are not aiming to improve on the
accuracy of previous calculations, indeed our estimates will be
considerably worse, being accurate to maybe 15\% at best.) Analytic
solutions also allow us to address, in more detail than is usually
possible, two questions:

\begin{itemize}
\item 
{\em Are the traditional UFB bounds either necessary or sufficient?}
\item
{\em Can $R$-parity violating models avoid metastability 
in the UFB direction?}
\end{itemize}

We will base our discussion on the Constrained MSSM (CMSSM)
which contains the usual $R$-parity invariant MSSM superpotential, 
\be
W_{MSSM}=h_U Q H_2 U^c + h_D Q H_1 D^c + h_E L H_1 E^c+\mu H_1 H_2,
\ee
and a degenerate pattern of supersymmetry breaking with universal
mass-squareds ($m_0^2$), trilinear couplings ($A_0$) and gaugino masses
($M_{1/2}$) at the GUT or Planck scale.  The notation is the same as in
Ref.\cite{waoi}. The extension to other patterns of symmetry breaking
is straightforward. In addition we will consider only the two most
important constraints on the CMSSM at low $\tan \beta $ which are those
due to Komatsu~\cite{komatsu} and Casas \etal~\cite{casas}. (A similar
analysis could be carried out on the other directions.) The
corresponding flat directions we will refer to as directions `K' and
`C' respectively.

Direction K is given by~\cite{komatsu} 
\ba
\label{komkom}
h_2^0             &=& -a^2 \mu/h_{D_j} \nonumber \\
\tilde{d}_{L_j}=\tilde{d}_{R_j} &=& a \mu/h_{D_j} \nonumber \\
\tilde{\nu}_i   &=& a  \sqrt{1+a^2} \mu/h_{D_j} 
\ea
for any pair of generations $i$, $j$, where $a$ parameterises distance
along the flat direction.  (This does not give the `fully optimised'
condition~\cite{ccb,casas}, but the difference will not be important
here.) The potential along this direction is $F$ and $D$-flat, and
depends only on the soft supersymmetry breaking terms;
\be
\label{softv}
V=\frac{\mu ^2}{h_{D_j}^2} a^2 (a^2 (m_2^2+m_{L_i}^2) + 
m_{L_i}^2+m_{d_j}^2+m_{Q_j}^2 ).
\ee 
At large values of $a$, therefore, the potential appears to be
unbounded unless $m_2^2 + m_{L_i}^2 >0$. Once radiative corrections are
taken into account (and the very weak assumption that $m_2^2 +
m_{L_i}^2 >0$ at the GUT scale made), one instead finds that certain
choices of soft supersymmetry breaking parameters cause the potential
to develop a minimum radiatively at some high scale which depends on
the weak scale parameters such as $\mu $.  In the CMSSM at the
quasi-fixed point for example the resulting traditional UFB bound is
$m_0\geqsim M_{1/2}$~\cite{ccb,casas}.

Direction C corresponds to,
\ba
h_2^0             &=& -a^2 \mu/h_{E_j} \nonumber \\
\tilde{e}_{L_j}=\tilde{e}_{R_j} &=& a \mu/h_{E_j} \nonumber \\
\tilde{\nu}_i   &=& a  \sqrt{1+a^2} \mu/h_{E_j} 
\ea
where $i\neq j$. The potential along this direction is 
\be
V=\frac{\mu ^2}{h_{E_j}^2} a^2 (a^2 (m_2^2+m_{L_i}^2) + 
m_{L_i}^2+m_{L_j}^2+m_{e_j}^2 ).
\ee
In both of these cases we work in the basis in which the relevant
Yukawa coupling is diagonal.  The bound from direction C is comparable
to that from K.

In the next section we address the cosmological aspects of these flat
directions, and argue that the answer to the first of our questions is
{\em no}. That the UFB bounds are not necessary is already well known;
the decay rate into a UFB direction at finite temperature is suppressed
by a large, temperature-dependent, barrier in the effective
potential~\cite{riotto}.  To show that it is not sufficient either, we
need a more detailed understanding of the potential which we obtain
through our analytic approximations.

Replacing the traditional UFB bound with a necessary condition would
require detailed knowledge of early cosmology which is, unfortunately,
lacking. We shall instead propose a truly sufficient condition which
arises from requiring that there is only one minimum in the
zero-temperature effective potential and that it is at the origin. By
`sufficient', we mean that we do not have to appeal to any cosmological
model (even the so called Standard one) to end up in the correct
minimum.

In the third section we develop an analytic treatment of both the
traditional and sufficient bounds. We explain why they are numerically
very close (and can in fact only be resolved with two-loop or better
accuracy).

In the fourth section we address the second of our questions.  It
arises because, as is the case for nearly all $F$ and $D$-flat
directions, the K and C directions correspond to (analytic)
gauge-invariant polynomials which are absent from the
superpotential~\cite{carlos}.  Stated in this way, it is clear that it
is discrete symmetries that lead to UFB problems/metastability. Of
course some discrete symmetry is required to prevent baryons decaying.
However, in models which violate $R$-parity, the most dangerous
directions can be lifted, although it is not clear {\em a priori}
whether it is possible to add all the required couplings without
violating experimental bounds. It turns out that the necessary
couplings {\em can} be added and destroy the dangerous minima if the
$R$-parity violation is just below experimental bounds.

\section{Vacuum decay into and out of flat directions}

Before getting to grips with detailed calculations of the bounds, we
will address the first question above which concerned the relevance of
the traditional UFB bound. To do this we need to find the decay rates
into and out of the dangerous K and C directions.  In this section we
estimate the false vacuum decay rate into the radiatively induced
minima at finite temperature.  We also consider what happens when the
new minimum which develops radiatively along the flat direction is not
the global minimum. In this case we find that the decay rate back to
the global minimum at the origin is also greatly suppressed. Hence we
will conclude that whether the new minimum is global or local (the
criterion of the traditional UFB bound) is irrelevant.  The only
relevant bound we can impose (apart from the need to avoid quantum
tunneling into the charge and colour breaking minimum which only
excludes very low values of $m_0$~\cite{riotto}) is a sufficient one --
that there be no minima at all along the UFB directions except the
physical one.

We first borrow some results from later sections to get some idea of
the shape of the potential along the $F$ and $D$ flat direction.
Figure 1 shows 
\be
\tilde{V}=  V h_b^2 /\mu^2 M_{1/2}^2
\ee
in the K direction (with $j=3$) in the CMSSM for the three values of
$m_0 $ indicated. It is plotted in the approximation that we can
neglect the bottom and tau Yukawa couplings and two-loop corrections in
the running of parameters. (A full two-loop numerical running produces
something similar.) We have taken $\mu = 500\gev $, $m_b (m_t)=2.5
\gev$ and $\alpha_3(m_t)=0.108$.  The potential is taken at the
quasi-fixed point so we don't have to worry about the value of $A_0$
but is otherwise typical.  As can be seen, a new minimum forms
radiatively when $m_0/M_{1/2}\leqsim 1.07$ and becomes global when
$m_0/M_{1/2}\leqsim 1.05$. The latter value corresponds to the
traditional UFB bound (modulo the approximations we are making here).
The minimum in the flat direction (which we shall denote by $\phi $)
forms when $a\equiv a_{min}=8.5$ so that $\phi_{min}
 = few \times 10^6 \gev $.  One fact which will prove extremely useful
when deriving the bounds analytically later is that it is the first
term in the potential (\ie $m_2^2+m_L^2$) which dictates the form and
position of the radiatively induced minimum. The depth of the minimum
is therefore $\sim m_W^2 \phi_{min}^2 = a_{min}^4 \mu^2 m_W^2 /h_b^2$.

Now consider the decay rate into the new minimum at finite
temperature.  In Ref.\cite{riotto} it was argued that the traditional
UFB bounds are not necessary because the decay rate into the UFB
directions is generally far too small.  The suppression occurs due to
the presence of a large, temperature (T) dependent barrier between the
false vacuum (where we are living if the UFB bound is violated) and the
true charge and/or colour breaking one.  This barrier is generically
present due to the flat direction, $\phi $, giving masses to fields to
which it couples. For $\phi \geqsim T$ the contribution to the
temperature dependent effective potential from a field, $i$, is
suppressed as $e^{-m_i(\phi)/T}$ (where $m_i$ is the mass it derives
from coupling to the flat direction) and the potential is the same as
that at zero temperature with a curvature of order $m_W^2$ coming from
the soft supersymmetry breaking terms (\cf Eqn.(\ref{softv})).  For
$\phi \sim 0$ the contribution to the effective potential is ${\cal{O}}
(-T^4) $. The escape point of the critical bubble is therefore of order
$\phi\sim T^2 /m_W$. Since the barrier height is of order $T^4$ the
decay rate is suppressed by a factor $e^{-S_3/T}$ where $S_3\sim
T^4/m_W^3 $, which is huge unless $T < m_W $. At zero temperature the
quantum tunneling rate is only large enough for vacuum decay to have
taken place within the lifetime of the universe for very low values of
$m_0$~\cite{riotto}.

The estimates of Ref.\cite{riotto} apply equally to the dangerous K and
C directions we are considering here.  Actually we can make a much more
accurate estimate of the action for the finite temperature decay when
$m_\phi \ll T\ll \phi_{min} $ as follows.  The one-loop finite
temperature contribution to the effective potential is
\be
V_T(\phi)={T^4\over 2\pi^2}\sum_i\pm n_i\int_0^\infty dq\  q^2{\rm
ln}\left[ 1\mp{\rm exp}\left(-\sqrt{q^2+m_i^2(\phi)/T^2}\right)\right],
\ee
where the upper sign is for bosons, the lower one for fermions, and the
$n_i$ are the corresponding degrees of freedom. The masses $m_i$ are
field dependent parameters which generally have a wide range of values
and increase roughly linearly with distance along the flat direction.
The potential to calculate the decay rate is normalised to zero at the
origin,
\be
\Delta V_T(\phi) = V_T(\phi)-V_T(0).
\ee 
The integral has the well known limits when $m_i \ll T$ and 
$m_i \gg T$;
\be
\Delta V_T(\phi)=
\left\{
\begin{array}{cc}
 {T^2\over 24}\sum\left[n_B\Delta m_B^2+{n_F\over
2}\Delta m_F^2\right] & \mbox{ ;}\nn\\
 \pi^2\bar g_*T^4/90 &\mbox{ ;}
\end{array}
\right.
\begin{array}{c}
 m_i(\phi) \ll T \nn\\
 m_i(\phi) \gg T
\end{array}
\ee
where $\Delta m^2\equiv m^2(\phi)-m^2(0)$ and 
\be
\bar g_* =n_B+(7/8)n_F
\ee
counts the difference between the degrees of freedom of the heavy
particles at that particular value of $\phi$ (\ie those for which $m_i
\gg T$) and at $\phi=0$.  Particles which are strongly coupled to the
flat direction cause a steep rise in the effective potential near the
origin.  The contribution near $\phi=0 $ can also be expressed as
$V(\phi)=(m_T^2-m_\phi^2) \phi^2/2 $ where $m_\phi^2\sim m_W^2$ is the
(negative) zero temperature mass-squared of the $F$ and $D$-flat
direction and
\be
m_T^2 = 
{T^2\over 24}\sum\left[n_B\lambda_B^2+{n_F\over
2} \lambda_F^2\right]
\ee
in which $\lambda_{B,F}$ stands for the coupling of the field to the
flat direction ($\sim g_3$,~$g_2$ or $h_t$ in this case).  (We take the
mass-squared to be constant (and always negative) for this estimate.)
The finite-temperature contribution to the effective potential of
particles which couple to the flat direction can therefore be
approximated by
\be 
V(\phi)=\frac{\kappa^2}{2} \phi^2 T^2 \Theta (\kappa T - \kappa' \phi)+
\left(\frac{\kappa^{\prime 2}}{2} T^4 + 
\frac{m^2 _\phi}{2} \left( \frac{\kappa^2}{\kappa^{\prime 2}} T^2 -
\phi^2\right)\right) \Theta ( \kappa' \phi - \kappa T),
\ee    
where we have defined 
\ba
\kappa^2 &=& (m_T^2 - m_\phi^2)/T^2 \nn\\
\kappa^{\prime 2} &=&  \pi^2\bar g_*/45.
\ea 
The first term is the $\phi \ll T$ limit, the second term is the $\phi
\gg T$ limit, and the last term is the zero temperature contribution
which is still present when $\phi \gg T$.  (The second $T^2$ term is
just to match at $\kappa'\phi= \kappa T$.) For the moment we are
assuming that all the particles couple to $\phi $ equally and one
should bear in mind that in the MSSM there are of course many different
couplings. To get the decay rate we find the 3-dimensional euclidean
action for a (spherically symmetric) critical bubble, $\phi(\rho)$
where $\rho $ is the radial coordinate. The bubble is a solution of the
equation~\cite{linde}

\be 
\label{crit}
\frac{\partial^2 \phi}{\partial \rho^2}+
\frac{2}{\rho}\frac{\partial\phi}{\partial\rho}
=\frac{\partial V}{\partial\phi},
\ee
with the boundary conditions
\ba
\phi (\infty)&=&0 \nonumber \\ 
\left.\frac{\partial \phi}{\partial\rho}\right|_0 &=& 0.
\ea
By matching $\phi $ and $\partial \phi/\partial \rho $ at $\kappa' \phi 
= \kappa T$, we find the solution
\be
\phi(\rho)= \left\{ \begin{array}{cc}
\frac{\kappa T}{\kappa' } \frac{\hat{\rho}}{\rho}
\frac{\sin \rho m_\phi }{\sin \hat{\rho} m_\phi} 
& \rho < \hat{\rho}\nonumber \\
& \nonumber \\
\frac{\kappa T}{\kappa' } \frac{\hat{\rho}}{\rho} 
e^{(\hat{\rho}-\rho) \kappa T} & \rho > \hat{\rho}
\end{array}
\right. 
\ee
where the radius of the bubble is given by 
\be
\hat{\rho}=\frac{1}{m}\left(\pi- \tan^{-1} (m_\phi/\kappa T)
\right) \approx \pi /m_\phi
\ee
and the escape point by 
\be
\phi(0)= \frac{\hat{\rho} m_\phi \kappa T }{\kappa' 
\sin(\hat{\rho} m_\phi )}
\approx \frac{\pi \kappa^2}{\kappa'} \frac{T^2}{m_\phi},  
\ee
where the approximations are for $\kappa T \gg m_\phi $. 
The three dimensional euclidean action is then given by 
\ba
S_3/T &=& 4\pi \int \rho^2 \dif \rho \left( \frac{1}{2} 
\left(\frac{\dif \phi}{\dif\rho}\right)^2
+ V(\phi) \right)\nn\\
&=&
 \frac{2 \pi^4 \kappa'^2  }{3} \frac{T^3}{m_{\phi}^3}
\left( 
1-\frac{1}{\pi}\tan^{-1}(m_\phi/\kappa T)\right)^3 \left( 1+
\frac{\kappa^4}{\kappa'^4} \left( \frac{m_\phi}{\kappa T}\right)^2 \right)
\nn\\
&\approx &
 \frac{2 \pi^6 \bar g_*  }{135} \frac{T^3}{m_{\phi}^3}.
\ea
This action is independent of the precise couplings to the flat
direction because we assumed that the temperature is high enough for it
to be dominated by the very large escape point.  Hence the action is
the same when there are many different couplings to the flat direction,
$\lambda_i$, provided that they are all either strong enough for
particle $i$ to freeze out well before the escape point or weak enough
for it to not contribute appreciably to
the effective potential. These two conditions can be written; 
\ba
\mbox{ either $\lambda_i$} &\gg & m_\phi/ T \nn\\
\mbox{ or $\lambda_i$} &\ll & m_\phi/\phi_{min}.
\ea

The criterion for the vacuum to have decayed within the lifetime of 
the universe is (see for Ref.~\cite{riotto} for example)
\be
S_3/T \leqsim 200.
\ee
For $T\gg m_{\phi}$ the action is indeed much more than $~200 T $ and
becomes larger for higher temperatures, from which we conclude that the
false vacuum is unlikely to have decayed within the lifetime of the
universe.  At $T\sim m_{\phi}$ we would need to take into account all
the field directions in order to estimate the rate as well as the depth
of the electroweak symmetry breaking minimum which we were able to
neglect above.  However from the above and Ref.\cite{riotto} it seems
unlikely that the false vacuum would decay within the lifetime of the
universe even in this case.

Now consider decay from the new minimum to the origin. This would be
relevant if either the new minimum were local (\eg if $1.05 M_{1/2} <
m_0 < 1.07 M_{1/2} $ in figure 1) or if it were global at zero
temperature but lifted by finite temperature corrections (of order
$T^4$ where $T\ll \phi_{min} $).  The barrier height (in figure 1 for
example) is governed by the first term in the zero temperature
potential, and is therefore of order $a_{min}^4 \mu^2 m_W^2 /h_b^2$,
and its width is of order $\phi_{min} = a_{min}^2 \mu /h_b$.  The
critical bubble is determined by Eq.(\ref{crit}) in which the RHS is of
order $(a_{min}^2 \mu /h_b) m_W^2 $.  We can obtain an estimate of the
action by scaling out the dependence on parameters using the
dimensionless variables $\tilde{\phi}=\phi h_b /(a_{min}^2\mu) $ and
$\tilde{\rho}=m_W \rho $.  At finite temperature a critical bubble has
a three dimensional euclidean action given by
\be
S_3=4\pi \frac{a_{min}^4 \mu^2 }{h_b^2 m_W} f
\ee
where $f$ is the integral, 
\be
f =\int_0^\infty \tilde{\rho}^2 \mbox{d} \tilde{\rho} \left(\frac{1}{2}
\left(\frac{\partial \tilde{\phi} }{\partial\tilde{\rho}}\right)^2+
V \frac{h_b^2 }{a_{min}^4 \mu^2 m_W^2}\right)
,
\ee
which we expect to be of order $1$.  The decay rate is therefore only
significant for large temperatures.  Decay out of the false vacuum (for
the potential of figure 1) within the lifetime of the universe would
require,
\be
200 T\geqsim 4 \pi \frac{a_{min}^4 \mu^2 }{h_b^2 m_W}\sim 5\times 10^{11} \gev.
\ee 
Of course our approximation breaks down long before this temperature is
reached since, when $T\gg \phi_{min} \sim 10^6 \gev$, the finite
temperature contributions to the effective potential (proportional to
$\phi^2 T^2 $ for $T \gg \phi $) will in any case destabilise the local
minimum.  The quantum tunneling rate is also suppressed; for
significant tunneling within the lifetime of the universe the $O(4)$
symmetric `bounce' solution~\cite{coleman} should have an action $S_4
\leqsim 400$. We estimate
\be
S_4 = 2\pi^2 \frac{a_{min}^2 \mu^2 }{h_b^2 m^2_W}f' \gg 400
\ee
where the integral 
\be
f =\int_0^\infty \tilde{\rho}^3 \mbox{d} \tilde{\rho} \left(\frac{1}{2}
\left(\frac{\partial \tilde{\phi} }{\partial\tilde{\rho}}\right)^2+
V \frac{h_b^2 }{a_{min}^4 \mu^2 m_W^2}\right)
,
\ee
which is again expected to be $\sim 1$.

With these two pieces of information to hand, it is clear that the
traditional UFB bound is irrelevant; {\em it can never be either
necessary or sufficient}. It is not necessary because if the universe
were trapped in a metastable minimum at the origin it would not have
decayed within its lifetime. It is not sufficient because had the
universe been trapped in a radiatively induced local minimum it would
also not have been able to decay.

Finally, we can address one of the suggestions in Ref.\cite{riotto} for
driving the universe to a metastable vacuum at the origin.  Some sort
of dynamical mechanism is probably required if such a scenario is to
avoid fine tuning. This mechanism could simply be a sufficiently high
temperature, $T_R$, in the early universe. Each field, $i$, then
contributes positive curvature to the potential for $ \lambda_i \phi <
T_R$.  (Other proposed mechanisms depend on the additional soft
supersymmetry breaking terms which can be generated during
inflation~\cite{hsquared}.) However the destabilisation of the
radiatively induced minimum can only be {\em guaranteed} if $T_R$ is
greater than the zero-temperature position of the radiatively induced
minimum, $\phi_{min}$. For the examples we are considering,
\be
T_R \geqsim few \times 10^6 \gev.
\ee 
However when, as we are assuming, supersymmetry is communicated
gravitationally there is a bound on the reheat temperature coming from
nucleosynthesis; if $T_R$ is too high, nucleosynthesis is disrupted by
the creation and subsequent late decay of gravitinos~\cite{subir}.
This translates into the bound,
\be
\label{grav}
T_R \leqsim 10^{9}\gev.
\ee
Thus, when $m_0$ is close to the traditional UFB bound, the reheat
temperature has to fall inside the window which is left.  Moreover,
this window closes when the traditional UFB bound is more strongly
violated. Figure 2, which we will derive shortly, shows the position of
extrema along direction K at the quasi-fixed point plotted against
$m_0^2/M_{1/2}^2$ (with the same choice of parameters as in figure 1).
To the right of the cross are only minima, to the left maxima. The
cross corresponds to the point of inflexion in the $m_0=1.14 M_{1/2}$
curve of figure 1.  As $m_0^2$ decreases below the traditional UFB
bound, the minimum deepens and $\phi_{min}$ grows rapidly. For low
enough $m_0^2$, $T_R$ has to violate the bound in Eq.(\ref{grav}) if it
is to destabilise the radiatively induced minimum.  (For the
quasi-fixed point example in figure 2, the window disappears when
$m_0\leqsim 0.7 M_{1/2} $.) If the reheat temperature is far below
$\phi_{min}$, then we have to ensure that the contribution to the
effective potential from the more weakly coupled fields (proportional
to $\lambda_i^2 \phi^2 T^2 $ with $\lambda_i\ll 1$) is enough to drive
$\phi $ to the origin, a complicated and model-dependent task.

\bigskip

To summarize, without a detailed knowledge of early cosmology, there is
no well-defined criterion for excluding regions of parameter space
based on the existence of deep minima along $F$ and $D$-flat
directions, apart from at very small values of $m_0$ where quantum
tunneling can take place within the lifetime of the universe.  In the
following sections therefore, as well as deriving the traditional UFB
bound analytically, we shall focus on a sufficient condition which is
better defined, namely that there are no minima other than the one at
the origin.  This condition ensures that the origin is reached from any
initial $\phi $ even at zero temperature.  The areas of parameter space
which satisfy this condition are therefore allowed regardless of
cosmology, and figure 1 suggests that, at least for the UFB directions
we are considering here, they may not be very different from those
allowed by the traditional UFB bound.  At the quasi-fixed point for
example, figure 1 gives the traditional bound to be $m_0\geqsim 1.05
M_{1/2}$ and the sufficient condition to be $m_0\geqsim 1.07 M_{1/2}$.
Thus, in the one-loop approximation we are making (and possibly to
two-loop order as well), the traditional UFB bound is indistinguishable
from the sufficient condition.

\section{Analytic derivation of `UFB' bounds}

\noindent In this section we develop an analytic treatment of both the
traditional bound and sufficient condition coming from the K and C
directions in the CMSSM at low $\tan \beta $.  These are the two most
restrictive constraints and were analysed numerically in
Ref.~\cite{casas}.  We first summarize the one-loop RGE solutions, and
then use them to obtain the bounds.

We shall obtain the bounds two ways in an order which may at first seem
a little odd; we first present an accurate and straightforward but
semi-graphical method, and then we present an entirely analytic method
that requires some approximations. The reason for this ordering (it is
after all more usual to first derive something approximately and then
do it accurately) is that the second method is far more powerful,
allowing us to produce bounds which are valid for all low $\tan\beta$.
It also gives us more insight into how the bounds depend on the unknown
parameters (such as $\mu $) and why they are numerically so close to
each other.  In fact we will conclude that (as anticipated in the
previous section) two-loop accuracy is required to resolve the
traditional bound and the sufficient condition.

To be able to do an analytic treatment, we will need to express the
mass-squared parameters appearing in the potential as a function of
scale (\ie not as a numerical integral).  This is possible if we work
to one-loop and in the approximation that we can neglect couplings
other than $\alpha_2$, $\alpha_3$ and $h_t$ in the running.  We first
give solutions for all the parameters in terms of the three quantities
which are {\em strongly} attracted to quasi-fixed points in the minimal
supersymmetric standard model (MSSM) (although other terms formally
have quasi-fixed points too).  The three parameters with quasi-fixed
points are \ba
R &\equiv & h_t^2/g_3^2 \nonumber \\
A_t &\equiv & A_{U_{33}} \nonumber \\
3 M^2 & \equiv & m_{U_{33}}^2 + 
m_{Q_{33}}^2 +m_{2}^2 .
\ea
When $h_t$ is high at the GUT or Planck scale (higher than 1 say),
these three parameters are completely determined at the weak scale 
irrespective of the pattern of supersymmetry breaking at the GUT scale;
\ba
R^{QFP} &=& 0.87 \nonumber \\
A_t^{QFP} &=& -1.60 M_{1/2} \nonumber \\
(M^2)^{QFP} &=& 1.83 M_{1/2}^2.
\ea
Their values govern the running of the MSSM at low $\tan \beta
$~\cite{hill}. (As pointed out in Refs.\cite{carena,waoi} and
references therein, various constrained versions of the MSSM lead to
many more quasi-fixed points in addition to these three ubiquitous
ones, and can lead to enhanced predictivity and reduced FCNCs at low
$\tan\beta$.)

For completeness we present all the solutions for the flavour diagonal
parameters here and for the moment leave in the $\alpha_1$ dependence;
defining
\ba
\tilde{\alpha}_i &=& \frac{\alpha_{i}}{\alpha_i|_0}\nn\\
\delta_i^{(n)} & = & \tilde{\alpha}^n_i-1 \nn\\
G &=& \frac{R}{R_0} \tilde{\alpha}_3^{-43/9} 
\tilde{\alpha}_2^{-3} \tilde{\alpha}_1^{-5/99}
\ea
where the $0$-subscript indicates values at the GUT scale, these are
\ba
A_{U_{ii}}-\smallfrac{1}{2}A_t &=& M_{1/2}\left(-\smallfrac{8}{9}\delta_3^{(1)}+
\smallfrac{3}{2} \delta_2^{(1)} +\smallfrac{13}{99} \delta_1^{(1)}\right) + 
(A_{U_{ii}}-\smallfrac{1}{2}A_t)|_0\nonumber \\
A_b-\smallfrac{1}{6}A_t &=& M_{1/2}\left(-\smallfrac{40}{27}\delta_3^{(1)}+
\smallfrac{5}{2} \delta_2^{(1)} +\smallfrac{29}{99} \delta_1^{(1)}\right) + 
(A_b-\smallfrac{1}{6}A_t)|_0\nonumber \\
A_{D_{ii}} &=& M_{1/2}\left(-\smallfrac{16}{9}\delta_3^{(1)}+
3 \delta_2^{(1)} +\smallfrac{7}{99} \delta_1^{(1)}\right) + 
(A_{D_{ii}})|_0\nonumber \\
A_{E_{\alpha \alpha}} &=& M_{1/2}\left(
3 \delta_2^{(1)} +\smallfrac{3}{11} \delta_1^{(1)}\right) + 
(A_{E_{\alpha \alpha}})|_0\nonumber \\
B-\smallfrac{1}{2}A_t &=& M_{1/2}\left(\smallfrac{16}{9}\delta_3^{(1)}+
\smallfrac{3}{2} \delta_2^{(1)} +\smallfrac{5}{66} \delta_1^{(1)}\right) + 
(B-\smallfrac{1}{2}A_t)|_0\nonumber\\
m_{U_{33}}^2-M^2 &=& M^2_{1/2}\left(\smallfrac{8}{27}\delta_3^{(2)}+
\delta_2^{(2)} -\smallfrac{1}{27} \delta_1^{(2)}\right) + 
(m^2_{U_{33}}-M^2)|_0\nonumber\\
m_{Q_{33}}^2-\smallfrac{1}{2}M^2 &=& M^2_{1/2}\left(\smallfrac{16}{27}\delta_3^{(2)}-
\delta_2^{(2)} +\smallfrac{5}{297} \delta_1^{(2)}\right) + 
(m^2_{Q_{33}}-\smallfrac{1}{2} M^2)|_0\nonumber\\
m_2^2-\smallfrac{3}{2}M^2 &=& M^2_{1/2}\left(-\smallfrac{8}{9}\delta_3^{(2)}+
\smallfrac{2}{99} \delta_1^{(2)}\right) + 
(m^2_2-\smallfrac{3}{2}M^2)|_0\nonumber\\
m_{L_{\alpha\alpha}}^2 &=& M^2_{1/2}\left(-\smallfrac{3}{2}\delta_2^{(2)}-
\smallfrac{1}{22} \delta_1^{(2)}\right) +
(m^2_{L_{\alpha\alpha}})|_0\nonumber\\
m_{1}^2 &=& M^2_{1/2}\left(-\smallfrac{3}{2}\delta_2^{(2)}-
\smallfrac{1}{22} \delta_1^{(2)}\right) +
(m^2_{1})|_0\nonumber\\
m_{U_{ii}}^2 &=&  
M^2_{1/2}\left(\smallfrac{8}{9}\delta_3^{(2)}-
\smallfrac{8}{99} \delta_1^{(2)}\right) +(m^2_{U_{ii}})|_0\nonumber\\
m_{Q_{ii}}^2 &=& 
M^2_{1/2}\left(\smallfrac{8}{9}\delta_3^{(2)}-\smallfrac{3}{2}\delta_2^{(2)}
-\smallfrac{1}{198} \delta_1^{(2)}\right) +(m^2_{Q_{ii}})|_0 \nonumber\\
m_{D_{\alpha\alpha}}^2 &=& 
M^2_{1/2}\left(\smallfrac{8}{9}\delta_3^{(2)}-
\smallfrac{2}{99} \delta_1^{(2)}\right) +(m^2_{D_{\alpha\alpha}})|_0
\nonumber\\
\mu &=& \mu |_0 G^{1/4}
\ea
where $i = 1,2$ and $\alpha = 1,2,3$.  
 
The solutions of the remaining three parameters $R$, $A_t$ and $M^2$,
can be expressed as functions of 
\be
r\equiv 1/\tilde{\alpha}_3=
1+\frac{6\alpha_0}{4\pi}\log \frac{\Lambda}{M_{GUT}},
\ee
so that
\be
\tilde{\alpha}_2=\frac{3}{4-r}\mbox{ ; } 
\tilde{\alpha}_1=\frac{5}{16-11 r}. 
\ee
Taking $\alpha_3(m_t)=0.108$ means that $0.37<r<1$
with $r=1$ corresponding to the GUT scale. We shall 
use $r^{-1}$ and $\tilde{\alpha_3}$ intechangeably.
If we define 
\ba
\Pi(r) &=& \tilde{\alpha}_3^{16/9}
\tilde{\alpha}_2^{-3}
\tilde{\alpha}_1^{-13/99}
\nn\\
\hat{J}&=& \frac{1}{r\Pi(r)}\int ^1_r \Pi(r') \dif r' 
\ea
then, in terms of
the GUT scale values ($R_0$,~$A_0$,~$m_0^2$) plus
the functions 
\ba
R^{QFP}&=& 1/2\hat{J}\nn\\
\rho &=& \frac{R}{R^{QFP}} \nn\\
\xi &=& \frac{1-r}{r\hat{J}} -1 
\ea
we find
\ba
\frac{1}{R} &=& \frac{1}{R^{QFP}}+\frac{1}{R_0 \Pi r} \nn\\
A_t &=& (1-\rho)A_0 + M_{1/2} \left( \rho \xi -
(1-r) \left(\frac{16}{9}\tilde{\alpha}_3 +
\tilde{\alpha}_2 +\frac{13}{45} \tilde{\alpha}_1\right)\right)\nn\\
M^2 &=&  (1-\rho )m_0^2 - \frac{1}{3}\rho K +\frac{2}{3} 
M_{1/2}^2 (1-r)\gamma
\ea
where 
\ba
K &=& (1-\rho)(\xi M_{1/2}-A_0)^2 - \xi^2 M_{1/2}^2 + M_{1/2}^2 (1-r) (\xi+1)
\left(\frac{16}{9}\tilde{\alpha}_3 +\tilde{\alpha}_2 +\frac{13}{45} \tilde{\alpha}_1\right)
\nn\\
\gamma &=& \frac{16}{9}\tilde{\alpha}_3 (1+\tilde{\alpha}_3 (1-r)/2)
+\tilde{\alpha}_2(1-\tilde{\alpha}_2 (1-r)/6) 
+\frac{13}{45} \tilde{\alpha}_1(1-11\tilde{\alpha}_1 (1-r)/10).
\ea
There is only one integral to do, $\hat{J}$, which we must 
now approximate by dropping the $\alpha_1 $ terms 
\be
\hat{J} r \Pi \approx -\frac{16479}{1540}+\frac{r^{-7/9}}{4620}(14080+36960 r-1680 r^2+77r^3).
\ee
This approximates the full integral to about 5\% in the 
region of interest (\ie where the minimum develops). 

To make contact with the usual parameterisation of the CMSSM we can
express the GUT scale parameter $R_0$ as a function of of the weak
scale parameter $\tan\beta $ for which we need to insert the running
top quark mass, $m_t(m_t)= 167 \gev$, and use
\be 
R= \frac{m_t^2}{4 \pi \alpha_3 v^2 \sin^2\beta }
\ee
where $v=174.1 \gev$. Inserting the solution for $R$ 
gives 
\ba
\label{r0}
R_0^{-1}&=&\Pi r_t \frac{4\pi\alpha_3v^2}{m_t^2}  
\left( 
\sin^2\beta-\sin^2\beta^{QFP}
 \right) \nn\\
&=& 6.34
\left( 
\sin^2\beta-\sin^2\beta^{QFP}
 \right) .
\ea
In this approximation $(\tan\beta)^{QFP}=1.89$. However the value we
infer for $R_0 $ is clearly highly dependent on its exact value and
hence is sensitive to two-loop corrections. In this sense $R_0$ is a
more natural parameter to use than $\tan\beta $; the latter is
equivalent to a very peculiar weighting of the GUT scale parameter, so
that mid-range values of $\tan \beta $ all correspond to roughly the
same minimum value of $R_0$. As a nod to convention, however, we will
present some of our results in terms of $\tan\beta $. \\

\bigskip 

Now let us use these expressions to examine the potential of
Eq.(\ref{softv}) in more detail for the case of the CMSSM in the $j=3$
direction. We first use a semi-graphical method to extract the
traditional bound and sufficient condition.

The mass-squared parameters run with scale, and the appropriate scale,
$\Lambda$, to evaluate them at is where the finite one-loop
contributions (which we are neglecting) to the effective potential are
small, $\Lambda\sim h_t \phi $ (assuming that the minimum occurs for
$a\gg 1$ which we shall verify in a moment).  This should correspond
roughly to the largest mass if we were to diagonalise the mass matrices
and evaluate the one-loop contribution completely.  $a$ can then be
expressed as a function of $r$,
\be
a^2(r)=b(r) \exp (c(r-1))
\ee
where
\ba
\label{central}
b(r) &=& \frac{M_{GUT}h_b}{h_t \mu}= 
\frac{2.85\times 10^{12}}{ h_t(r)/1.1} 
\left(
\frac{M_{GUT}}{2\times 10^{16} \gev}
\right)
\left(
\frac{m_b(r)}{2.5\gev }
\right)
\left(
\frac{200 \gev}{\mu(r)}
\right)\nonumber \\
c &=& 4 \pi /6\alpha_3|_0 = 52.4
\ea
where the central values are those we shall adopt for the scale 
$\Lambda=m_t$ or 
$r=r_t$. We can solve for $b(r)$ in the 
same manner as above, and find 
\ba
b(r) &=&  b|_0 
\left(\frac{R}{R_0}\right)^{-2/3} 
\tilde{\alpha}_3^{14/27}
\tilde{\alpha}_2^{-1/2}
\tilde{\alpha}_1^{-7/594}\nn\\
&\approx &
b(r_t) \left(\frac{R(r)}{R(r_t)}\right)^{-2/3} 
\left(\frac{\alpha_3(r) \alpha_2(r_t)}{\alpha_3(r_t)\alpha_2(r)}\right)^{1/2}
\ea

Denoting the position of the radiative minimum by $\phi_{min}$, 
the traditional bound is saturated by 
\be
\left. V\right|_{\phi_{min}}=\left.\frac{\partial V}{\partial\phi}
\right|_{\phi_{min}}=0
\ee
and the sufficient condition is given by
\be
\left.\frac{\partial V}{\partial\phi}\right|_{\phi_{min}}
=\left.\frac{\partial^2 V}{\partial\phi^2}
\right|_{\phi_{min}}=0,
\ee
and so we can work with 
\be
\tilde{V}=  V h_b^2 /\mu^2 M_{1/2}^2
\ee
which depends only on $b(r)$. (This means that for the 
sufficient condition we are neglecting the {\em running} of $h_t$ which 
is relatively small in this region.) The potential becomes
\ba
\tilde{V}(r,m_0)&=&\frac{a^2(r)}{M_{1/2}^2}
\left[ 
a^2(r) (\smallfrac{3}{2} M^2+\smallfrac{1}{2}m_0^2
-M_{1/2}^2  (
\smallfrac{8}{9}\delta_3^{(2)}+
\smallfrac{3}{2}\delta_2^{(2)}+
\smallfrac{5}{198}\delta_1^{(2)}
))\right.
\nonumber\\
&&+\left.
\smallfrac{1}{2} M^2+\smallfrac{5}{2}m_0^2
+M_{1/2}^2  (
\smallfrac{40}{27}\delta_3^{(2)}-
\smallfrac{5}{2}\delta_2^{(2)}-
\smallfrac{29}{594}\delta_1^{(2)}
)\right]
\ea
Figure 1 shows $\tilde{V}$ for $m_0=1.03 M_{1/2}$,~$1.05 M_{1/2}$, and
$1.07 M_{1/2}$, for the central values of parameters we used in
Eqn.(\ref{central}) and for the quasi-fixed point where $R_0^{-1}=0$
and where there is no dependence on $A_0$ in the mass-squareds.  The
minima of $V$ are at $\phi_{min}=few\times 10^6 \gev $.  At the
quasi-fixed point the traditional bound is saturated with a minimum
corresponding to $a_{min}=a(0.5427)=8.5$, justifying our assumption of
$a^2\gg 1$ in direction K.

As can be seen, for these values of $m_0$ the minimum is being rapidly
lifted, with $m_0=1.05 M_{1/2}$ saturating the traditional UFB bound.
From this figure we have already
anticipated one of our main conclusions; radiatively induced minima
along UFB directions are lifted very quickly by changes in the
parameters ($m_0$ in this case).  This happens because the negative
$a(r)^4 $ term in the potential balances the positive $ a(r)^2$ one
when the traditional UFB bound is saturated, so that the height of the
minimum is varying rapidly.

Ideally one would like to be able to locate the extrema for a given
$m_0$, $R_0$ and $A_0$, see for what value of $m_0$ there exists a
minimum away from the origin, and hence derive the traditional UFB
bound.  To make the problem tractable analytically we can use the fact
that there is a one-to-two correspondence between $m_0$ and the
location of the extrema, $r=r_{min}$.  Consider again the K direction,
with a minimum at $\phi_{min}=a_{min}^2\mu^2/h_b^2 $ corresponding to
\be
r_{min}=1+\frac{6\alpha(M_{GUT})}{4\pi }\log \left(\frac{a_{min}^2}{b}\right),
\ee
and denote the value of $m_0$ corresponding to 
an extremum at $r=r_{min}$ by $m_0(r_{min})$. Solving for $m^2_0(r_{min})$ 
is relatively simple since $V(r,m_0)$ depends only linearly upon it; 
\ba
\label{mo}
{m^2_0(r_{min})}&=&\frac{M_{1/2}^2}{d(a^2+\smallfrac{5}{2})}\left[
-(a^2+\smallfrac{1}{3})R(r_{min})r_{min}^{-1}(3 M^2 + A_t^2)
- d(a^2+\smallfrac{1}{6})M^2
 \right. \nonumber\\
&& \mbox{\hspace{2cm}}+\smallfrac{32}{9} \tilde{\alpha}^3_3 + 
2 (a^2+1)\tilde{\alpha}^3_2 + 
\left( 
\smallfrac{2}{5}a^2 + \smallfrac{14}{15}\right) \tilde{\alpha}^3_1   
\nonumber\\
&& \mbox{\hspace{1cm}}\left. 
+d \left( 
\smallfrac{16}{9}a^2-\frac{40}{27}\right)\delta^{(2)}_3
+d \left( 
3 a^2 + \smallfrac{5}{2}\right)\delta^{(2)}_2 
+d \left( 
\smallfrac{5}{99}a^2 + \smallfrac{29}{594}\right) \delta^{(2)}_1
\right]
\ea 
where $d(r)$ encompasses the running of $a^2$;
\ba
d(r)&=&\frac{1}{a^2}\frac{d a^2}{d r}= \frac{d \log b}{dr}+c\nonumber\\
&=& \frac{1}{2}\tilde{\alpha}_2+
 \frac{1}{6}\tilde{\alpha}_1-\frac{4}{3 r}R+c.
\ea
Eq.(\ref{mo}) is implicit in that, away from the fixed point, $M^2 $
itself depends upon $m_0^2$. However, since the dependence is (by
dimensions) also linear the full (but unwieldy) expression can easily
be obtained from the above.

For a given value of $A_0$ and $R_0$ (or equivalently $A_0$ and
$\tan\beta $) we can now find the sufficient condition from
Eq.(\ref{mo}).  First consider the quasi-fixed point.  Figure 2 shows
$m_0^2(r_{min})/M_{1/2}^2$ plotted against $r_{min}$.  Values of
$r_{min}$ to the left of the cross are maxima, to the right, minima of
the potential for a given value of $m_0$.  This function has a maximum
when $m_0^2=1.14 M_{1/2}^2$ so that the only minimum when $m_0>1.07
M_{1/2} $ is at the origin, as we saw in figure 1.  Close to the
quasi-fixed point the sufficient condition can be approximated by
expanding in $1/R_0$ and by noting that, from the analytic solutions,
they can be expressed as
\be
\label{approx}
\frac{m_0^2}{M_{1/2}^2} > c_0+\left( c_1 + 
c_2 \frac{A_0}{M_{1/2}} + c_3 \frac{A_0^2}{M_{1/2}^2}\right)
/R_0 + 
{\cal{O}}(1/R_0^2).
\ee
By matching at the quasi-fixed point and at $R_0=5$ and
$A_0=(-1,~0,~1)$, we find that $c_{0..3}$=~1.14,~-1.69,~-0.63,~0.30
respectively.  These numbers approximate the bound (in the one-loop
approximation) to better than 5\% for $R_0\geqsim 4$ (corresponding to
$h_t(M_{GUT})>1.4 $) and $|A_0| \leqsim M_{1/2}$ and of course much
better for larger $R_0$.  We can also find an `absolute' condition by
completing the square in Eq.(\ref{approx}); values of $m_0$ which
satisfy the sufficient condition in direction K must also satisfy
\be
\frac{m_0^2}{M_{1/2}^2} > c_0+(c_1-c_2^2/4 c_3) /R_0 = 1.1-2.0/R_0.
\ee
If $m_0^2 $ is below this bound a minimum is radiatively generated
whatever the value of $A_0$.

We can now find the traditional UFB bound graphically by plotting the
height of the extrema, $V(r_{min},m_0(r_{min}))$, as shown in figure 3.
At the quasi-fixed point, the minimum becomes negative at
$r_{min}=0.5427$, giving the bound $m_0\geqsim m_0(0.5427) = 1.05
M_{1/2}$.  Again by matching at the quasi-fixed point and at $R_0=5$
and $A_0=(-1,~0,~1)$, we find the coefficients
$c_{0..3}$=~1.10,~-1.67,~-0.63,~0.30.  Thus the sufficent condition is
as expected virtually indistinguishable from the traditional bound.

Applying the same treatment to direction C at low $\tan\beta $ (which
was studied numerically in Ref.\cite{casas}), we find
$c_{0..3}$=~1.15,~-1.54,~-0.57,~0.26 for the traditional UFB bound and
for the sufficient condition we find
$c_{0..3}$=~1.17,~-1.56,~-0.57,~0.26.  (For this direction the
definition of $a^2$ the same, but with $b(r_t)=1.9\times 10^{12}$ for
$m_{\tau}(r_t)=1.7\gev $.) These bounds are in agreement with
Ref.\cite{casas} given our neglect of two-loop contributions to the
running of the gauge and Yukawa couplings.  Ref.\cite{casas} finds
$m_0\geqsim M_{1/2}$ at the quasi-fixed point (which should be compared
with our result above of $m_0> 1.05 M_{1/2}$).

Finally let us consider the accuracy of the approximations we made.  We
saw that radiatively induced minima along UFB directions are lifted
very quickly by changes in $m_0$. This means that the bounds are less
sensitive to our lack of knowledge about the parameter $b$ and in fact
depend on it only logarithmically as we shall see shortly.  (In fact
the error involved is of the same order as that we associate with
neglecting the finite one-loop corrections.) We can also see that the
bound increases with $b$ as we would expect since increasing $b$
increases the negative contribution to the potential from $m_2^2$. This
is in accord with Ref.\cite{casas} where the effect of increasing the
`unification' scale to the Planck scale was examined.  We also see that
the bound increases with $m_b$ (which is why we chose $j=3$) and
decreases with $\mu $. The variation of the bound with $A_0$ is
quadratic becoming stronger at large $\tan\beta $ with a minimum close
to $A_0=0$ at positive $A_0$. \\

\bigskip

Although the above method is accurate for a given choice of $A_0$ and
$\tan\beta$ (given the other approximations we have made), it is
cumbersome and in particular is not well suited to finding bounds at
smaller values of $R_0 $.  In addition we would like to understand why
the sufficient condition is numerically so close to the traditional
bound.  Because of this we now turn to a completely analytic derivation
of the same bounds.

For this we can use the fact that the minimum appears at a scale where
the first term in the potential is driven negative by $m_2^2 $.  So,
considering direction K, if we define $A(r)=
(m_2^2+m_{L_i}^2)/M_{1/2}^2 $ and $ B(r)=
(m_{L_i}^2+m_{d_j}^2+m_{Q_j}^2)/M_{1/2}^2 $, the minimum of the
potential is at a scale close to the point which we shall call $r_p$
where $A$ becomes negative, $A(r_p)=0$. We define two additional
scales; $r_d\equiv$ the position of the minimum when the traditional
UFB bound is saturated (\ie when the minimum is degenerate with the
origin), and $r_c\equiv$ the position of the point of inflexion when
the sufficient condition is satisfied.  We first determine these scales
in terms of $r_p$ as follows.  It is convenient to work with
\be 
\hat{V}=\frac{h_t^2}{M_{1/2}^2 M_{GUT}^2} V=
\left(e^{2 c(r-1)} A + e^{c (r-1)} B/b \right),
\ee
(Note that instead of $M_{GUT}$ we could have used any scale, 
$\Lambda_X$, as long as $c=4 \pi /6 \alpha_3|_{\Lambda_X} $.) 
Since the scales $r_p$ will appear 
algebraically in the determination of the bounds 
we can sacrifice a little accuracy in 
their determination, and approximate the potential close 
to $r_p$ by developing $A(r)$;
\be
\hat{V}(r)\approx 
(r-r_p) \dot{A}(r_p) e^{2 c(r-1)} + e^{c(r-1)} B(r_p)/b(r_p) .
\ee
Because at this point $A$ is running much faster than $B$, $b$ and 
$h_t$, we are going to neglect the running of $B$, $b$ and $h_t$ 
close to $r_p$. For the bounds we will need 
\ba
\dot{\hat{V}}(r) &\approx &(1+2 c(r-r_p)) \dot{A} e^{2c(r-1)} +  e^{c(r-1)} 
c B/b 
\nn\\
\ddot{\hat{V}}(r) &\approx & 4c(1+c(r-r_p)) \dot{A} e^{2c(r-1)} + 
 e^{c(r-1)} c^2 B/b.
\ea
For the traditional UFB bound, ($\hat{V}(r_d)=\dot{\hat{V}}(r_d)=0$),
we find
\ba
\label{rpd}
r_p &=& 1+\frac{1}{c}\ln \left( \frac{e c B}{\dot{A}b}\right) \nn\\
r_d &=& r_p-1/c.
\ea
Likewise, for the sufficient condition, ($\ddot{\hat{V}}(r_c)=
\dot{\hat{V}}(r_c)=0$),
we find
\ba
\label{rpc}
r_p &=& 1+\frac{1}{c}\ln \left( \frac{e^{3/2} c B}{2 \dot{A}b}\right) \nn\\
r_c &=& r_p-3/2c.
\ea
Since $c$ is large (\cf Eq.(\ref{central})) the minima are very close
to $r_p$ (as was the case for the example shown in figure 1). The value
of $r_p$ in each case is determined using the expression for $B$ above
and for $\dot A$ which follows straightforwardly from the
renormalisation group equations~\cite{inoue},
\be
\dot A=\frac{R}{r}(3 M^2+A_t^2) -2  
\tilde{\alpha}_2^3-\frac{2}{5}
\tilde{\alpha}_1^3.
\ee
A comparison of the expressions for 
$B$ and $\dot A$ together with $A(r_p)=0$ gives 
\be
\dot A/B\approx R/r
\ee
to a good approximation independently of $m_0/M_{1/2}$.  This allows
for a sufficiently accurate determination of $r_p$ as the dependence on
the parameters above is only logarithmic. For relatively large
$R_0\geqsim 1$ it is well approximated by using the quasi-fixed values
of parameters in the logarithm
\be
r_p=0.019 \log (1.67+m_0^2/M_{1/2}^2 ) +\left\{ 
\begin{array}{cc}
0.544 & \mbox{traditional UFB}\nn\\
0.540 & \mbox{sufficient }
\end{array}
\right.
\ee
The values of $r_p$ are those which must satisfy 
\be
\label{arp}
A(r_p)=0=\smallfrac{3}{2} M^2+\smallfrac{1}{2}m_0^2
-M_{1/2}^2  (
\smallfrac{8}{9}\delta_3^{(2)}+
\smallfrac{3}{2}\delta_2^{(2)}+
\smallfrac{5}{198}\delta_1^{(2)}
),
\ee
and solving this gives the bounds on $m_0^2$ we are looking for. 
We find the approximate expression for the bound 
\be
\label{mapprox}
\frac{m_0^2}{M_{1/2}^2}
=
\frac{4\rho_p-\rho_p^2-1.67 + \rho_p (1-\rho_p)\alpha}
{4-3\rho_p}
\ee
where 
\ba
\frac{1}{\rho_p}
&=&
1+
\left.
\frac{R^{QFP}}{R_0\Pi r}\right|_{r_p}
\approx 
1+\frac{1}{3 R_0}
\nn\\
\alpha &\approx & \frac{A_0^2}{M_{1/2}^2}
-2 \frac{A_0}{M_{1/2}}.
\ea
We again see from Eq.(\ref{mapprox}) that the bounds are always
quadratic in $A_0$ (except for the very weak $A_0$ dependence in the
determination of $r_p$). We also see that the bounds are much more
restrictive for negative $A_0$ and have a minimum at $A\approx
M_{1/2}$.

As an example of the usefulness of this method we show in figure 4 the
`absolute' bound on $m_0$ in this direction for all $\tan\beta < 15 $
where our RGE solutions are accurate to one-loop order.  For values of
$m_0$ below this bound there is a radiatively induced minimum whatever
the value of $A_0$.  Replacing $\rho_p$ in Eq.(\ref{mapprox}) with
$\tan\beta$ gives a nice approximation to this diagram.  (Since the
diagram goes well away from the quasi-fixed point we found $r_p$ by
solving the first equations of Eqs.(\ref{rpd},\ref{rpc}) fully rather
than using the large $R_0$ approximation above.) The plateau in figure
4 is a result of the mid-range values of $\tan\beta$ all corresponding to
roughly the same value of $R_0$.

To compare this approximation with the previous method, we find the
coefficients for the expansion near the quasi-fixed point of
$c_{0..3}$=~1.13,~-1.67,~-0.63,~0.30 and
$c_{0..3}$=~1.09,~-1.65,~-0.63,~0.30 for the sufficient condition and
traditional UFB bounds respectively.

We can now see why the traditional UFB bound and the sufficient
condition are so close. The separation between the scales $r_d$ and
$r_c$ at which the minima form can only be of order ${\cal{O}}(1/c)$.
When we solve for $A(r_p)=0$ the values of $r_p$ differ by a factor
$e^{1/2}/2 =0.82 $ appearing in the logarithm. Thus we would have to
know $b$ and in particular $\mu $ to better than 18\% (\ie to
two-loops) before we could distinguish the two bounds. On the other
hand the parameter $r_p$ depends only logarithmically on $b$ but
appears algebraically when we determine the bounds -- hence the bounds
on $m_0$ can only depend logarithmically on $b$.

\section{Lifting flat directions with $R$-parity violation.}

In this section we will see {\em how much} the flat directions can be
lifted if we choose `baryon parity' rather than $R$-parity to prevent
proton decay~\cite{herbi}.

As we stated in the introduction, violating $R$-parity can lessen UFB
problems because there are less gauge invariants. The role of gauge
invariants in flat directions was discussed in Ref.\cite{carlos}.  To
clarify the role of the $R$-parity violating couplings let us briefly
summarize that work. If we represent the scalar component of a generic
superfield, $k$, as $z_k$, then the potential contains $D^2$ and $F^2$
terms where
\be
D^A={\overline{z}}_k T^A_{kl} z_l 
\ee
and $T^A$ are the matrices of the generators of the group. In addition,
gauge invariance of any analytic function $I(z_k)$ requires that
\be
I_k T^A_{kl} z_l =0
\ee
where the subscript on the $I$ implies differentiation by $z_k$.  Thus,
there exists a $D$-flat direction corresponding to every
\be
\label{inv}
I_k|_{\langle z\rangle } = C {\langle 
{\overline{z}}_k\rangle },
\ee
where $C$ is a complex constant. The inverse, \ie that every flat
direction corresponds to an invariant satisfying Eq. (\ref{inv}), is a
less trivial property.  If the invariants in question do not appear in
the superpotential, then the $F^2$ can often be made to cancel.  

If we didn't already know the UFB directions in the MSSM, then we could
have used Eq.(\ref{inv}) to do it methodically (which is extremely useful
for more complicated flat directions).  For the K directions, the
relevant invariants are 
\ba
W &=& H_1 ( h_{D_{j}} Q_jD_j + \mu H_2) \nn\\
I &=& L_i ( Q_jD_j + \mu' H_2)
\ea
where $W$ is a term in the superpotential while $I$ is not allowed if
one assumes $R$-parity. The K direction in Eq. ({\ref{komkom}) is found
as the solution of Eq.(\ref{inv}) for $I$ with the coefficient $\mu'$
being chosen to cancel $F_{H_1}$, $\mu'=-a^{2}\mu/h_{D_j}$.  Similarly, the
C direction can be constructed from the  $R$-parity violating invariants
\be
L_i L_j E_j \mbox{ ; } H_2 L_i
\ee 
with $i\neq j$.

Notice that the absence of a particular gauge invariant operator from
the superpotential is not enough {\em per se} to ensure $F$-flatness
unless no pair of superfields in the gauge invariant operator appear
together in any term of the superpotential. For example $Q_j$ and $D_j$
both appear together in the Yukawa couplings of the MSSM and this is
why we had to construct our flat direction by the conjunction of two
invariants in order to cancel $F_{H_1}$.  An example of a single gauge
invariant which corresponds to an $F$ and $D$-flat direction is the
other $R$-parity violating operator,
\be
\epsilon U D D .
\ee
This flat direction (although it has been discussed at some length in
the literature) is usually safe in models of gravitationally
communicated supersymmetry breaking.

Baryon parity allows all of the invariant operators above corresponding
to the K and C directions, and so can potentially lift all the flat
directions we have been discussing. Since the K and C directions with
$j=3$ result in comparable bounds on parameter space, we will assume
that all the directions should be lifted in order to significantly
change the conditions on parameters seen for example in figure 4. There
are 5 flat directions in total, corresponding to $i=1,2,3$ in the K
directions and $i=1,2$ in the C directions.  In this section we will
attempt to lift all five, whilst satisfying experimental constraints on
$R$-parity violating operators.

We begin by adding to the superpotential the additional operators 
\be
W_{B}
=\lambda_{ ijk}L_i L_j E_k + \lambda'_{ijk} L_i Q_j D_k .
\ee 
We do not consider the $R$-parity violating bilinear operators, $\mu_i
L_i H_2 $, since the dimensionful coefficient $\mu_i$ must be much less
than $ m_W$ and so their contribution to the potential at large field
values will be very small. Along the K and C directions the
contributions to the potential from the $F^2$ terms are
\ba
\label{fterms}
V_{F-terms} &= & \sum_m \frac{a^4 \mu^4}{h_b^4} 
\left(
    |\lambda'_{m33}|^2 
+ (1+a^2) |\lambda'_{im3}|^2 
+ (1+a^2) |\lambda'_{i3m}|^2 
\right)
\nonumber \\
V_{F-terms} &= & \sum_m 2 \frac{a^4 \mu^4}{h_\tau^4} 
\left(
    |\lambda_{m33}|^2 
+ (1+a^2) |\lambda_{im3}|^2 
+ (1+a^2) |\lambda_{i3m}|^2 
\right) 
\ea
respectively. In addition one can expect contributions from new
$R$-parity violating trilinear terms. Since the sign of the field (\ie
the sign of $a$) has not yet been determined, the potential will be
minimised when they are negative;
\ba
V_{A-terms} &\sim & -|A_{\lambda'}\lambda'_{i33}\mu^3 | 
\frac{|a|^3 \sqrt{1+a^2}  }{h_b^3} \nn\\
V_{A-terms} &\sim & -|A_{\lambda} \lambda_{i33}\mu^3 | 
\frac{|a|^3 \sqrt{1+a^2}  }{h_\tau^3}
\ea
respectively. 
The $F$-terms must overcome both these terms 
and the radiatively induced minimum at $\phi_{min}\approx a_{min}^2 \mu/h_b $
if they are successfully to lift the flat direction. 

We now discuss how big the new couplings have to be, generically, to
lift the flat directions. Let us begin by making a rough estimate in
which we say that the new $R$-parity violating $F$-terms have to `fill'
the radiatively induced minimum; \ie we are neglecting the fact that
the new terms change the position of the minimum.  (We shall see
shortly that this estimate is surprisingly good.) Since the depth of
the minimum is $\sim a_{min}^4 \mu^2 m_W^2 /h_b^2 $, the $R$-parity
violating couplings (generically denoted by $\lambda $ and $\lambda' $)
must satisfy
\ba
\label{lam}
\lambda \geqsim \frac{h_\tau m_W }{a_{min} \mu}
\approx 0.004 \frac{m_W}{\mu}\nn\\
\lambda' \geqsim \frac{h_b m_W }{a_{min} \mu}\approx 
0.004 \frac{m_W}{\mu}.
\ea
The couplings $\lambda $ and $\lambda'$ are understood to be evaluated
at the scale of the minimum.  If some of the flat directions are lifted
by the $a^4$ terms in Eq.(\ref{fterms}) then the $\lambda $ and
$\lambda' $ must be on the edge of experimental detection unless $\mu $
is very large~\cite{herbi}. We therefore concentrate on the $a^6 $
terms and the condition above.  Assuming that, as in canonical
gravitationally communicated supersymmetry breaking, $A_\lambda \sim
A_{\lambda'}\sim m_W$ and that all of the $R$-parity violating
couplings the same order of magnitude, the trilinear terms are
dominated by the $F$-terms
if Eq.(\ref{lam}) is satisified and $a_{min}\gg 1$.

We can refine our estimate substantially by adopting the approach of
the last section. If $R$-parity violating terms are to make flat
directions safe then we should, to be consistent, insist that there are
no minima at all except the Standard-Model-like minimum for {\em any}
value of parameters. Defining
\be
\hat{\lambda}=\frac{\lambda \mu}{h_\tau M_{1/2}},
\ee
(and similar for $\lambda' $) the form of the potential is now 
\be 
\tilde{V}= V h_\tau^2 /\mu^2M_{1/2}^2 
=a^4 A + a^2 B + a^6 \hat{\lambda}^2 ,
\ee
where $a(r)$,~$A(r)$ and $B(r)$ are as before.  We now find the
sufficient condition -- \ie a condition which ensures that the
$R$-parity violation has destroyed all minima along the flat direction.
The sufficient condition is then
\be
\label{triv}
\left.
\frac{d^2{\tilde{V}}}{da^2}\right|_{r_c}=
\left.
\frac{d{\tilde{V}}}{da}\right|_{r_c}=0
\ee
where $A$ and $B$ are to be evaluated at the scale $r_c$ which we have
yet to find. The determination of $r_c$ is simplified once we realise
that all three terms are essential in the lifting of the minimum. The
$a^6 \hat{\lambda}^2$ term simply pushes the (extremely deep) minimum in
the potential to low energy scales until it runs up against something
positive -- in this case the $a^2 B$ term. Thus a suitable criterion
for the determination of $r_c$ is simply that $a^2 A(r_c)+B(r_c)=0$.
Once we have $r_c$ we solve Eq.(\ref{triv}) treating $A$ and $B$ as
constants and obtain the critical value of $\hat{\lambda}$. (The
accuracy is extremely good because, in contrast to the determination of
the UFB bounds, $A$ and $B$ are running very slowly at $r_c$).  The
value of $\hat{\lambda}$ is dependent on $m_0$, but if we want to add
enough $R$-parity violation to remove the UFB problem entirely, we
should do it at $m_0=0$ where the minimum is deepest.  We therefore
choose $m_0=0$ in our evaluation of $r_c$ and also (for simplicity)
$A_0=0$. We find critical values
\be
0.76 >\hat{\lambda}_{crit} > 0.50 
\ee
with the upper value corresponding to $\tan\beta=1.8$ and the lower
value to $\tan\beta =15 $.  (It drops to a plateau in $\tan\beta $ in
much the same way as figure (4).) The revised estimate for the
$R$-parity violation required to remove {\em all} UFB bounds is
therefore
\ba
\label{lam3}
\lambda \geqsim \frac{0.76 h_\tau M_{1/2} }{ \mu}
\approx 0.017 \frac{M_{1/2}}{\mu}\nn\\
\lambda' \geqsim \frac{0.26 h_b M_{1/2} }{\mu}\approx 
0.009 \frac{M_{1/2}}{\mu}.
\ea

Given that it is the $a^6$ terms which lift the flat directions, what
set of $R$-parity violating couplings do we need?  For the two C
directions we can get away with having only one large coupling,
$\lambda_{123}$. This coupling can satisfy Eq.(\ref{lam3}) without
violating current experimental bounds assuming that $\mu $ is not very
much smaller than $M_{1/2}$; charged current universality requires that
\be 
\lambda_{123}<0.05 \left( \frac{m_{\tilde{e}_R}}{100 \gev}\right).
\ee
For the K directions, we need at least three relatively large
$\lambda'$ couplings. In addition if we were, for example, to choose
them to be $\lambda'_{113}$, $\lambda'_{213}$ and $\lambda'_{313}$,
then it is clear that we could just rotate the leptons into a basis
where this is equivalent to just one coupling and we would still be
left with two $F$-flat directions.  We need to choose either
$\lambda'_{1a3}, \lambda'_{2b3}, \lambda'_{3c3}\neq 0$ or
$\lambda'_{13a}, \lambda'_{23b}, \lambda'_{33c}\neq 0$, where $a,b,c$
are all different (or any choice of three of these where the last two
indices are all different).  As an example, we choose $\lambda'_{113},
\lambda'_{223}, \lambda'_{333}\neq 0$. The last two of these couplings
are hardly constrained at all, and can easily satisfy experimental
bounds~\cite{herbi} and Eq.(\ref{lam}) simultaneously.  The first one
however appears in quite restrictive bounds involving products of
itself with one other $\lambda' $ coupling. It is also constrained by
charged current universality to be \be \label{lam2} \lambda'_{113}<0.02
\left( \frac{m_{\tilde{d}_R}}{100 \gev}\right).  \ee Applying
Eq.(\ref{lam3}) to the product bounds in Ref.\cite{herbi} results in new
bounds for some of the other $R$-parity couplings.  Assuming that the
squark masses are $100 \gev ,$ for consistency with  Ref.\cite{herbi},
they are\\

\begin{center}
\begin{tabular}{|c|c|c|}
\hline & & \\
coupling $\leqsim $ & `UFB' limit $\times (\mu/M_{1/2})$ 
& experimental limit \\
\hline & & \\
$\lambda'_{111} $ & 0.0008 & 0.00035 \\
$\lambda'_{112} $ &  0.04 & 0.02 \\
$\lambda'_{121} $ & 0.03 & 0.035 \\
$\lambda'_{122} $ & 0.04 & 0.02 \\
$\lambda'_{131} $ & $9.0\times 10^{-6}$ & 0.035 \\
$\lambda'_{132} $ & 0.17 & 0.33 \\
$\lambda'_{213} $ & $6.0\times 10^{-6}$ & 0.09 \\
\hline 
\end{tabular}\\
\end{center}
\bigskip

\noindent When $\mu \approx M_{1/2} $ we see that most of the 
new bounds are comparable to the current experimental ones. The
bounds on $\lambda'_{132}$ and  $\lambda'_{121}$ are actually 
more restrictive and those on  $\lambda'_{131}$ and  $\lambda'_{213}$ 
much more restrictive. 

To summarize, in minimal supersymmetry with constrained (degenerate)
supersymmetry breaking, the addition of $R$-parity violation just below
the current experimental bounds (\cf Eq.(\ref{lam3})) is enough to lift
dangerous flat directions and destroy local minima. For values of
$\tan\beta $ which are away well from the quasi-fixed point the
magnitude of the $R$-parity violating couplings required to do this
drops rapidly by a factor of $\sim 2$. 

\section{Conclusions}

Using approximate analytic solutions to the RGEs of the MSSM, we have
examined the formation of minima in $F$ and $D$ flat directions. We
have given arguments which suggest that the traditional UFB bounds are
not meaningful since, even if a radiatively induced minimum is not
global, the decay rate back to the physical vacuum is very small.  We
have rederived the traditional UFB bounds analytically, and shown that
they are numerically very close to satisfying a more meaningful
sufficient condition -- that there be no radiatively induced minima
along the flat direction at all.  The sufficient condition was derived
for all $\tan \beta < 15 $.

In the literature the question of dangerous flat direction is usually
tackled by either making an implicit or at least not very detailed
appeal to cosmology, or by completely excluding regions of parameter
space where the physical vacuum is metastable -- a condition which, as
we have seen, is probably not even relevant.  The unbiased approach we
have been advocating is to locate in the parameter space regions where
some (unknown) cosmological input is needed to save the model. The
`traditional' UFB bound does {\em not} do this since it is still
possible to get caught in a charge and colour breaking minimum even if
it not the global one (and hence satisfies the `traditional' UFB
bound). The sufficient condition we gave is only slightly more
restrictive than the `traditional' bound but is in our opinion the only
possible relevant condition and is very different in spirit.

We also examined the effect of $R$-parity violation, and found that a
small amount of $R$-parity violation (consistent with but just below
current experimental limits) is able to lift the most dangerous flat
directions and remove radiatively induced minima.
In this way $R$-parity removes the metastability problems which 
plague the MSSM, and rescues some models which could be worth
saving. One example is the MSSM at the quasi-fixed point which has much
less dependence on initial parameters (as we have seen) and hence is
more predictive~\cite{carena,waoi}. This model is likely to be
severely constrained by a combination of UFB bounds and higgs bounds
coming from LEP2~\cite{carena,sabel}. A second example is the dilaton
breaking scenario which has degenaracy in the supersymetry breaking
terms (at tree level) and hence naturally suppressed flavour changing
neutral currents. This model was excluded because of UFB bounds in
Ref.\cite{casas2}.

\section{Acknowledgements}
We would like to thank Ben Allanach and Sacha Davidson for discussions
and for a critical reading of the manuscript and also Toby Falk for 
additional comments.

\newpage

\begin{figure}
\vspace*{-4in}
\hspace*{-2.0in}
\epsfysize=12in
\epsffile{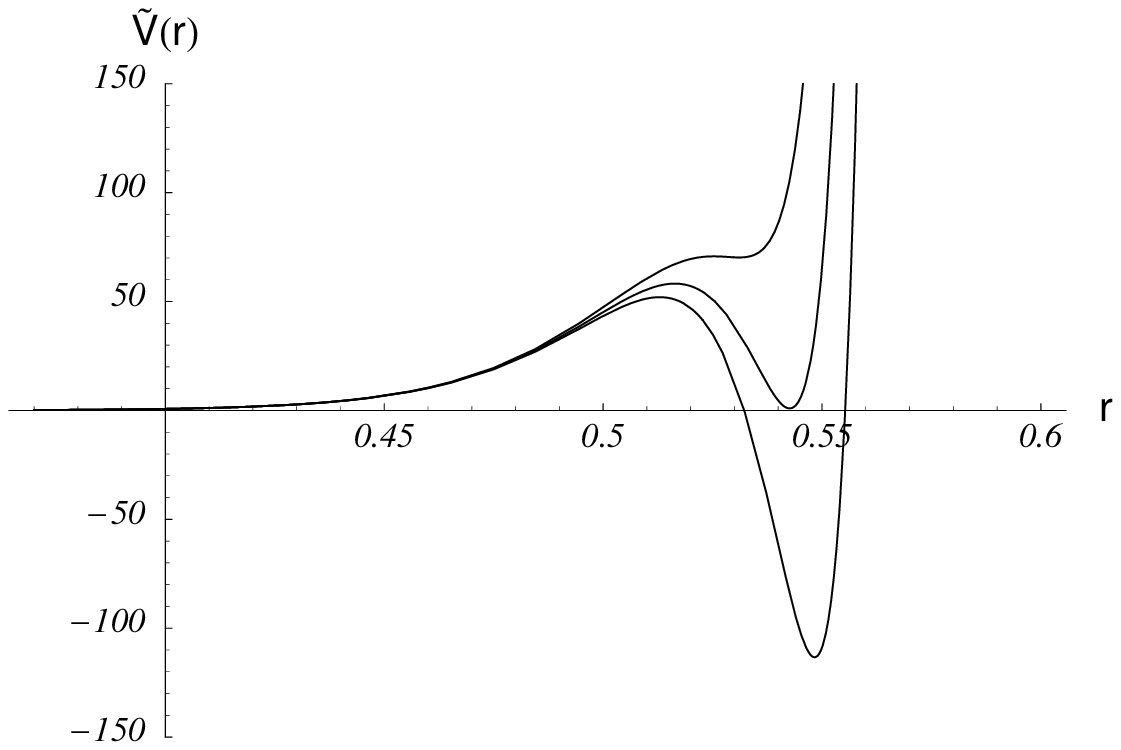}
\vspace{-2in}
\caption{Potential along the flat `K' direction, $\phi$, 
vs field values (parameterized by $r=1+\frac{6\alpha(M_{GUT})}{4\pi}\log 
\left(\frac{h_t \phi}{M_{GUT}}\right)$) 
for $m_0/M_{1/2}=$~1.03,~1.05,~1.07.} 
\end{figure}

\newpage

\begin{figure}
\vspace*{-4in}
\hspace*{-2.0in}
\epsfysize=12in
\epsffile{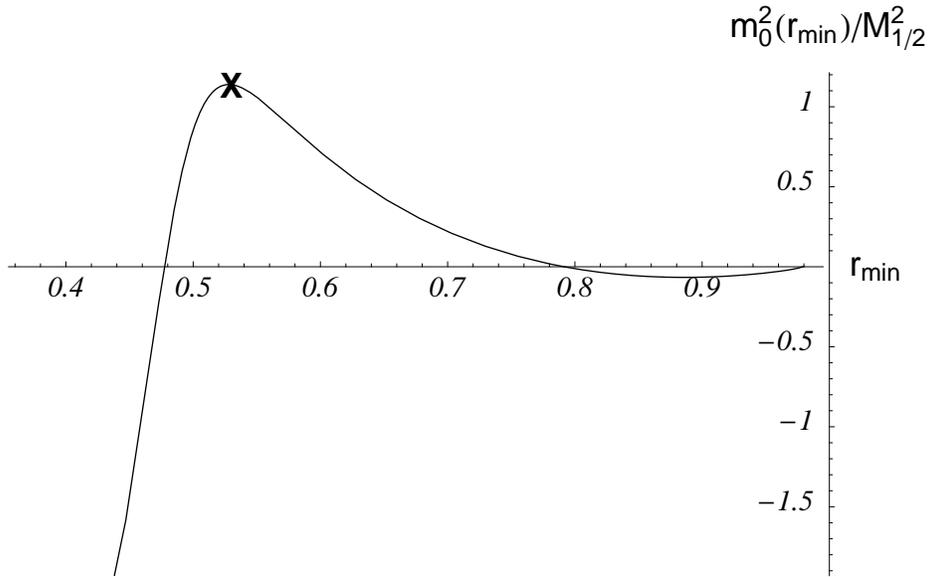}
\vspace{-2in}
\caption{The value of $m_0^2$ corresponding to an extremum appearing 
at that field value (parameterized by $r$). Each value of $m_0$ 
generates one maximum (to the left of the cross) and one minimum (to the 
right). This example is at the quasi-fixed point.} 
\vspace*{2in}
\end{figure}
\vspace{2in}

\newpage

\begin{figure}
\vspace*{-4in}
\hspace*{-2.0in}
\epsfysize=12in
\epsffile{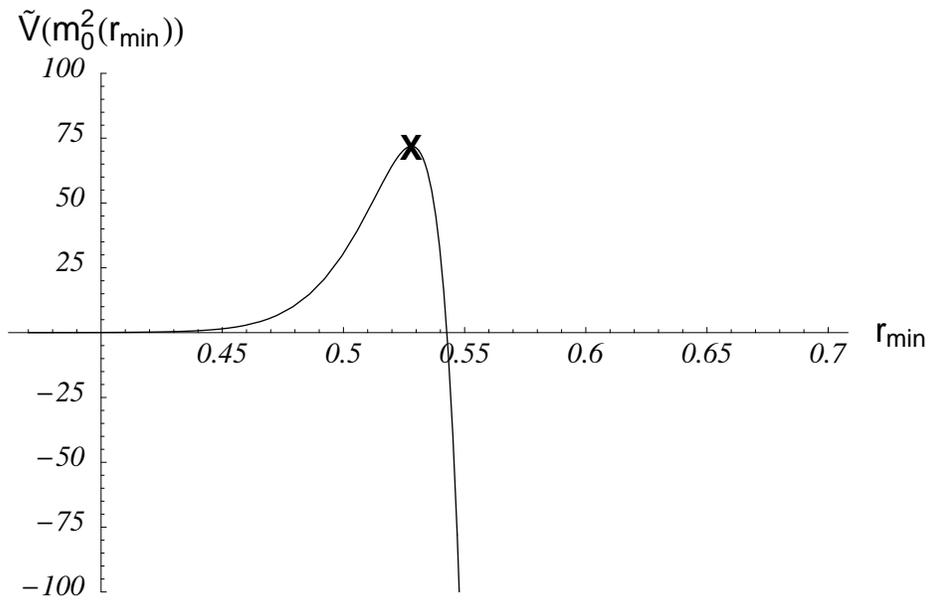}
\vspace{-3in}
\caption{The height of the extrema vs their field value (parameterized by
$r$). This example is at the quasi-fixed point and corresponds to 
substituting the $m^2_0$ values in figure 2 into the potential.} 
\end{figure}

\newpage

\begin{figure}
\vspace*{-4in}
\hspace*{-2.0in}
\epsfysize=12in
\epsffile{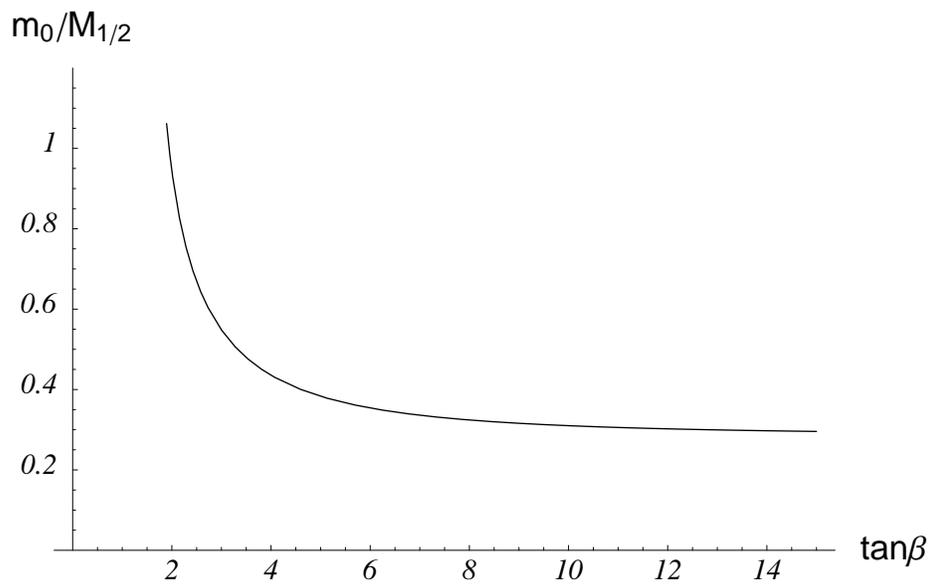}
\vspace{-3in}
\caption{The sufficient condition for $\tan\beta <15 $ and 
for any value of $A_0$. A minimum is formed radiatively 
when $m_0$ is below the line. } 
\end{figure}

\end{document}